\documentclass[10pt]{iopart}

\usepackage{setspace}
\usepackage{graphicx}
\usepackage{lmodern}
\usepackage{float}
\usepackage{upgreek}
\usepackage{amssymb}
\usepackage{color}
\usepackage[normalem]{ulem}
\usepackage{cancel}
\usepackage{miller}
\usepackage{siunitx}

\DeclareSIUnit{\angstrom}{\textup{\AA}}

\definecolor{jg_color}{cmyk}{1, 0.3, 0, 0}
\definecolor{ns_color}{cmyk}{0, 0.6, 1, 0}
\definecolor{ch_color}{cmyk}{0.5, 0, 1, 0}
\definecolor{eg_color}{cmyk}{0.7, 1, 0, 0}
\definecolor{darkred}{cmyk}{0,1,1,0.2}

\begin{document}
	
	\title[Uniaxial-stress-induced magnetic transitions in the triangular-lattice antiferromagnet PdCrO$_2$]{Uniaxial-Stress-Induced Magnetic Transitions in the
Triangular-Lattice Antiferromagnet PdCrO$_2$}

	\author{Nina Stilkerich$^{1,2}$, Tobias Ritschel$^1$,  Hilary M. L. Noad$^2$,  Richard Waite$^3$, Dmitry Khalyavin$^3$, Kousuke Ishida$^4$,  Pascal Manuel$^3$, Fabio
Orlandi$^3$, Seunghyun Khim$^2$, Elena Gati$^{2,1}$, Andrew P. Mackenzie$^{2,5}$, Jochen Geck$^{1,6*}$ and Clifford W. Hicks$^{7*}$}
	
	\address{$^1$ Institute of Solid State and Materials Physics, Technische Universität Dresden, 01069 Dresden, Germany}
	\address{$^2$ Max Planck Institute for Chemical Physics of Solids, Nöthnitzer Str. 40, 01187 Dresden, Germany}
	\address{$^3$ ISIS Facility, Rutherford Appleton Laboratory, Chilton, Didcot, OX11 0QX, United Kingdom}
	\address{$^4$ Institute for Materials Research, Tohoku University, Sendai 980-8577, Japan}
	\address{$^5$ Scottish Universities Physics Alliance (SUPA), School of Physics and Astronomy, University of St. Andrews, St. Andrews KY16 9SS, United Kingdom}
	\address{$^6$ W\"{u}rzburg-Dresden Cluster of Excellence ctd.qmat, Technische Universit\"{a}t Dresden, 01062 Dresden, Germany}
	\address{$^7$ School of Physics and Astronomy, University of Birmingham, Birmingham B15 2TT, United Kingdom}
	\address{$^*$ authors to whom any correspondence should be addressed}
	
	\ead{jochen.geck@tu-dresden.de, c.hicks.1@bham.ac.uk}
	
	\vspace{10pt}
	\begin{indented}
		\item[]March 2026
	\end{indented}

	\begin{abstract} 
Uniaxial stress is a promising method to tune magnetic frustration, allowing its effects to be studied in a precise way. In this work, uniaxial stress is applied to the
triangular-lattice antiferromagnet PdCrO$_2$. The Cr-Cr magnetic interaction is very sensitive to interatomic separation, so laboratory-achievable stress can induce substantial
changes in magnetic structure. Results from three types of measurement are presented: X-ray diffraction, the stress-strain relationship, and neutron diffraction. The combined
data show that the elastic moduli of PdCrO$_2$ are strongly affected by stress-induced changes in magnetic structure. A new, first-order stress-induced magnetic transition is
observed, at which the lattice constant shrinks by 0.21\%. The lattice stiffens dramatically across this transition: the Young's modulus increases by $\approx 80$~GPa, and the
Poisson ratio falls from $\approx 1$ to $\approx 0.4$. This stiffening indicates that the magnetic order ``locks,'' that is, becomes insensitive to lattice strain. This locking
might occur because the new stress-induced magnetic order nests the Fermi surface of the Pd sheets. Other frustrated magnets, including candidate spin liquids, may show similarly
strong coupling between magnetic and elastic degrees of freedom.
	\end{abstract}
	
	%
	\vspace{2pc}
	
	
	%
	\submitto{\RPP}
	%
	%
	\ioptwocol

	\section{Introduction}

The simplest example of magnetic frustration is antiferromagnetic interaction on a triangular lattice. If the spins are Heisenberg spins and nearest-neighbour interaction
dominates, then the ground state on an isotropic triangular lattice is a $120^\circ$ spin spiral, for both classical and $S=1/2$ systems~\cite{8806Huse}. The effect of introducing
anisotropy --- stronger or weaker interaction for bonds along a selected direction --- is a well-studied theoretical problem~\cite{9906Weihong, 0908Tocchio, 2106Szasz}. An example
of an experimental approach to this problem is provided by triangular-lattice organic charge-transfer salts. The nearly-isotropic compound $\kappa$-(BEDT-TTF)$_2$Cu$_2$(CN)$_2$ is
a candidate spin liquid~\cite{0309Shimizu, 1201Jeschke}, while the more anisotropic compound $\kappa$-(BEDT-TTF)$_2$Cu[N(CN)$_2$]Cl is an antiferromagnet with a N\'{e}el
temperature of 27~K. Applying uniaxial stress to the former, to increase its magnetic anisotropy, causes transition into a new magnetic state, that may be static antiferromagnetic
order~\cite{2508Lieberich}.

\begin{figure}[ptb]
		\includegraphics[width=86mm]{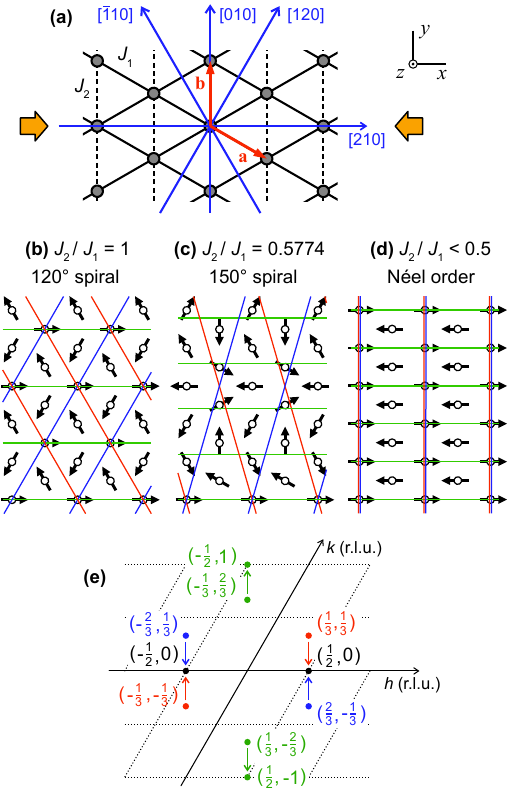}
		\caption{\label{fig_magOrderHypothesis} Lattice directions and hypothesised spin configurations under strain. (a) Schematic of a triangular lattice under uniaxial
stress applied along the $[210]$ lattice direction. $J_1$ is the exchange energy across the bonds that get compressed (solid lines), and $J_2$ across those that get stretched
(dashed lines). The blue lines are high-symmetry lattice directions. The $x$, $y$, and $z$ directions, used in discussion of strain, are also shown. (b) 120$^\circ$ spin spiral,
obtained when $J_2 = J_1$. The coloured lines are magnetic Bragg planes. (c) 150$^\circ$ spin spiral, obtained for $J_2 = 0.5774J_1$ (neglecting all longer-range interactions). (d)
N\'{e}el order. (e) Motion of the magnetic reflections as the magnetic order evolves from the 120$^\circ$ spin spiral to N\'{e}el order. The $h$ and $k$ axes are aligned with
reciprocal lattice vectors.  The colours correspond to those of the real-space Bragg planes indicated in panels b-d.}
	\end{figure}

The goal of the measurements reported here is to probe how far uniaxial-stress tuning of a frustrated magnet can be taken, and to explore coupling between magnetic order and
lattice elasticity. The system studied here, PdCrO$_2$, is a classical triangular-lattice antiferromagnet, with $S = 3/2$. To see what might happen, consider a triangular lattice
placed under uniaxial stress along the $[210]$ direction, as illustrated in Fig.~\ref{fig_magOrderHypothesis}(a). Bonds along two of the bond directions get compressed equally, and
along the third direction, stretched. Denote as $J_1$ the exchange energy across the bonds that get compressed, and $J_2$ that along those that get stretched. If the spins are
classical Heisenberg spins and all interactions beyond nearest-neighbour are negligible, then it is a well-known but so far untested prediction that at $J_2/J_1 = 0.5$ there
should be a transition from spiral to N\'{e}el order, in which the spin-spin rotation angle is $180^\circ$~\cite{5906Villain, 5906Yoshimori}.

The magnetic order of PdCrO$_2$ below $T_\text{N} = 37.5$~K is almost exactly a $120^\circ$ spin spiral. For edge-sharing CrO$_2$ octahedra, the exchange energy increases as
interatomic separation is reduced~\cite{7005Mochida}, so we can anticipate $J_2 < J_1$. In both experimental and theoretical studies, the nearest-neighbour interaction in PdCrO$_2$
is found to be far stronger than the next-nearest-neighbour interaction~\cite{1807Le, 2011Komleva}, so a focus on nearest-neighbour interactions, as outlined above, is a
reasonable starting point. To tune $J_2/J_1$ from $\approx 1$ in unstressed PdCrO$_2$ to $\sim 0.5$ would seem to be a technically challenging goal, but the sensitivity of
magnetic interaction in chromium oxides to interatomic spacing is very high~\cite{7005Mochida, 1303Hardy}.

In PdCrO$_2$, the magnetic CrO$_2$ layers are separated by metallic Pd sheets. The Fermi surface of the Pd sheets has a rounded hexagonal shape, and a low effective
mass~\cite{1507Hicks}. The magnetic and metallic sub-systems interact observably: the magnetic order reconstructs the Fermi surface of the Pd sheets~\cite{1310Ok, 1401Noh,
1507Hicks, 2002Sunko}, and onset of magnetic order changes the Hall conductivity, a subject of considerable discussion~\cite{1009Takatsu, 1512Daou, 2405Jeon}. 

In unstressed PdCrO$_2$, the vector chirality of the spin spiral --- the spin rotation direction on moving from one Cr site to the next --- alternates from layer to
layer~\cite{0903Takatsu, 1403Takatsu}. This means that two propagation vectors are required to describe the magnetic order in unstressed PdCrO$_2$. Under uniaxial compression along
the $[210]$ direction, there is a first-order transition to a state in which the vector chirality is the same in all layers~\cite{2112Sun}. We will refer to this transition as the
double-single-$q$ transition, and it is apparent in the data below, but not the main subject of this paper.

The main subject of this paper is what happens at larger stress. Key lattice directions that will be referred to throughout this paper are illustrated in
Fig.~\ref{fig_magOrderHypothesis}(a), and the progression from a $120^\circ$ spin spiral to N\'{e}el order in a nearest-neighbour-interaction-only model is illustrated in
Figs.~\ref{fig_magOrderHypothesis}(b-d). The propagation vector of an incommensurate spin spiral can and in general should change continuously as the lattice is strained, but
the N\'{e}el state, if it can be obtained, is expected to be rigid, that is, insensitive to further compression of the lattice.  Modeling presented below shows that such a
transition could have a large effect on elastic moduli. Therefore, we look for new magnetic transitions using two probes of elastic moduli: X-ray diffraction (XRD), through which
the differential Poisson ratio is obtained, and the stress-strain relationship, to probe the Young's modulus. Both of these methods have only recently been applied to quantum
materials~\cite{2111Sanchez,2310Noad, RV2411Hicks}. The data reveal large variation in the Poisson ratio and Young's modulus, associated with changes in magnetic structure. A
new transition is observed, across which the lattice stiffens dramatically. The modeling shows that such stiffening is expected from transition to a rigid magnetic order.  Neutron
diffraction data show that the new magnetic phase is indeed much less sensitive to strain than the magnetic orders at lower stress, but also that it is not the anticipated N\'{e}el
phase.

\section{Methods}

\subsection{Samples}

	PdCrO$_2$ single crystals were grown from PdCrO$_2$ polycrystals, synthesized using Pd, PdCl$_2$ and LiCrO$_2$, in a NaCl flux \cite{1000Takatsu} with liquid transport
growth \cite{1700Yan}. Four of the five samples studied here were sculpted using a Xe-ion focused ion beam (FIB) into a bowtie shape that concentrates stress~\cite{2012Ikhlas}, and
the fifth was mechanically cut and polished into a straight bar. The samples are listed in Table \ref{tab_samples}. As described elsewhere~\cite{1406Hicks}, for application of
uniaxial stress the sample ends are embedded in epoxy. The volumes in Table~\ref{tab_samples} refer to the central, exposed portion of the sample.

	\begin{table}[htbp]
		\centering
		\caption{List of the measured samples. $\times$ denotes samples that were shaped with a FIB.\vspace{1mm}}
		\begin{tabular}{c|c|c|c}
			\hline
			sample&technique&sculpted&volume (mm$^3$)\\
			\hline
			A&XRD									&$\times$	&3.6e-3\\ 
			B&$\sigma$-$\varepsilon$&$\times$	&4.8e-3\\
			C&$\sigma$-$\varepsilon$&$\times$	&1.2e-3\\
			D&neutron							&					&0.06\\ 
			E&neutron							&$\times$		&0.03\\ 
		\end{tabular}
		\label{tab_samples}
	\end{table}

\subsection{X-ray diffraction}

\begin{figure}[ptb]
		\includegraphics[width=86mm]{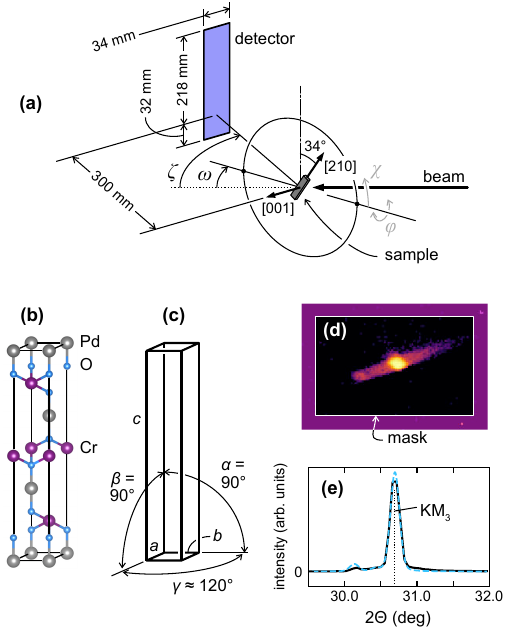}
		\caption{\label{fig_setup} Experimental information on XRD measurement under strain. (a) Setup for the X-ray diffraction measurement with detector angle $\zeta$,
cryostat rotation $\omega$, sample rotation $\varphi$, and sample tilt $\chi$ labelled. (b) Unit cell of PdCrO$_2$; this cell is of the obverse type. (c) A parametrisation of this
cell. (d) Sum of area detector images over varying $\omega$, focused on the $2\bar{2}\bar{5}$ reflection. (e) Azimuthally integrated intensity of the $2 \bar{2} \bar{5}$ reflection
(black line), along with a fit (blue dashed line).}
	\end{figure}

	X-ray diffraction measurements were performed on sample A using a custom-built, in-house 4-circle diffractometer equipped with a Pilatus CdTe 300K-W area detector. For the
measurements described below, the molybdenum KM$_3$ emission line ($\lambda = 0.6323$~\AA{}) was used as the X-ray source. For most of our measurements the sample rotation
$\varphi$ and tilt $\chi$ [see Fig.~\ref{fig_setup}(a)] were kept fixed, and only the cryostat rotation $\omega$ and the detector angle $\zeta$ were scanned. The sample was mounted
in a CS220T uniaxial stress cell from Razorbill Instruments, with the \hkl[210] direction at an angle of $34^\circ$ from the vertical. Measurements were done at $T = 14$ and 45~K,
using a helium pulse-tube cryostat with optical windows.
	
From the position of a pixel on the area detector and the detector angle $\zeta$, the corresponding scattering angle $2 \Theta$ --- the angle between the incident beam and
diffracted beam reaching that pixel --- can be determined. Extraction of $2 \Theta$ from one reflection is illustrated in Fig.~\ref{fig_setup}(d,e). The intensity of the
$2\bar{2}\bar{5}$ reflection, summed over multiple cryostat angles $\omega$, is shown in Fig.~\ref{fig_setup}(d). The intensity within a mask around the reflection was integrated
azimuthally~\cite{2502Kieffer}, yielding the intensity versus $2\Theta$ graph shown in Fig.~\ref{fig_setup}(e). This integrated intensity was then fitted using a three-peak model
representing the energy spectrum of the incident beam. The scattering angle $2\Theta$ was taken as the centre position of the main peak, corresponding to the KM$_3$ emission line.

Under uniaxial stress along the $[210]$ direction, the space group R$\bar{3}$m of unstressed PdCrO$_2$ is reduced to C2/m--- of the three mirror planes containing the $c$-axis,
only one is retained in the strained structure. The lattice parameters $a$, $b$, and $c$, and the unit cell angle $\gamma$ [see Fig.~\ref{fig_setup}(c)] of the strained structure
were determined at each applied stress by fitting the scattering angles 2$\Theta$ of selected reflections. Under the applied stress, we expect the unit cell angles $\alpha$ and
$\beta$ to remain at $90^\circ$, and the angle gamma to deviate only slightly from $120^\circ$.  Eleven reflections were selected for fitting. The rhombohedral space group
R$\bar{3}$m of PdCrO$_2$ allows for obverse and reverse structural domains--- ABC versus ACB stacking; the unit cell shown in Fig.~\ref{fig_setup}(b) is of the obverse type. The
eleven selected reflections were all from obverse structural domain(s). The obverse reflections were brighter and yielded lower-noise fitting than the reverse reflections,
indicating that a majority of the probed sample volume was of the obverse structure type.  The expression used for fitting was
	\begin{equation}
\label{eq_BraggAngle}
		\sin \Theta = \frac{\lambda}{2 \sin \gamma} \sqrt{\frac{h^2}{a^2} + \frac{k^2}{b^2} + \frac{2hk\cos \gamma}{ab} + \frac{l^2 \sin^2 \gamma}{c^2}},
	\end{equation}
where $h$, $k$, and $l$ are the reflection indices. This expression applies for $\alpha = \beta = 90^\circ$. We note that the fit results were averaged over two choices for the
lattice vectors $\mathbf{a}$ and $\mathbf{b}$: those illustrated in Fig.~\ref{fig_magOrderHypothesis}(a), and rotated counter-clockwise by $120^\circ$. These choices yield
different reflection indices, but are equivalent with respect to the stress axis.

	As a test of the accuracy of the fitted lattice constants, sample A was measured again after it had been fractured under tension--- the sample then consisted of two stubs
that could be held out of contact, guaranteeing that the applied stress was zero. In this case, we expect $a=b$. (The spontaneous lattice distortion associated with rotational
symmetry breaking is extremely small, well below the resolution of our XRD measurements~\cite{2507Stilkerich}.) Analysis of these zero-stress data yielded $a = 2.9139 \pm
0.0014$~\AA{} and $b = 2.9126 \pm 0.0012$~\AA{} (at $T = 14$~K), which agree within error and average to $a_0 = 2.9132 \pm 0.0009$~\AA{}.
	
	The measured lattice constants were converted to strain as follows. The lattice constants $a$ and $b$ are given by 
	\begin{eqnarray}
		\label{eq_a}
		a = a_0 \sqrt{\left(1 + \varepsilon_{xx}\right)^2 \sin^2 \gamma + \left(1 + \varepsilon_{yy}\right)^2 \cos^2 \gamma }, \\
		\label{eq_b}
		b = a_0 \times \left(1 + \varepsilon_{yy}\right),
	\end{eqnarray}
where $\varepsilon_{xx}$ is the strain along the $[210]$ direction, and $\varepsilon_{yy}$ that along $[010]$. Linearizing about $\gamma = 2\pi/3$ yields
	\begin{eqnarray}
		\label{eq_xstrain}
		\varepsilon_{xx} = \frac{4}{3}\frac{a}{a_0} - \frac{1}{3}\frac{b}{a_0} - 1, \\
		\label{eq_ystrain}
		\varepsilon_{yy} = \frac{b}{a_0} - 1.
	\end{eqnarray}
Eqs.~\ref{eq_xstrain} and~\ref{eq_ystrain} were used to obtain $\varepsilon_{xx}$ and $\varepsilon_{yy}$ from $a$ and $b$. The differential Poisson ratio is given by $\nu \equiv
-d\varepsilon_{yy}/d\varepsilon_{xx}$.

	\subsection{Stress-strain measurements}
	
Stress-strain data were collected on samples B and C using home-built uniaxial stress cells similar to that reported in Ref.~\cite{1902Barber}. These cells incorporate capacitive
sensors of both the force $F$ and displacement $D$ applied to the sample; information on conversion from the measured capacitances to $F$ and $D$ is provided in the Appendix. To
aid conversion to stress and strain, the samples were milled (using an Xe-ion focused ion beam) into bowtie shapes, with a narrow neck between two wide end tabs. This procedure is
detailed in Refs.~\cite{2012Ikhlas, 2310Noad}. The goal is to obtain a sharp crossover between the neck, which experiences high stress, and the end tabs, where the stress remains
low. In this way, the sample can be approximated as two springs in series: a Hooke's-law spring with spring constant $k_{anchor}$ representing the end regions, and a spring with an
arbitrary (though monotonic) stress-strain relationship representing the neck. Under this approximation, the relation between $D$ and strain $\varepsilon_{xx}$ is given by
	\begin{equation}
		\label{eq_twoSpring}
		D = \varepsilon_{xx}(\sigma_{xx})l_\text{neck} + \frac{F}{k_\text{anchor}}.
	\end{equation}
$\sigma_{xx} = F/A$ (where $A$ is the cross-sectional area of the neck) is the applied stress, and $l_\text{neck}$ is the length of the neck.  $\varepsilon_{xx}$ calculated using
Eq.~\ref{eq_twoSpring} most accurately reproduces the actual sample strain when $l_\text{neck}$ is taken as the length of the straight portion of the neck, excluding the fillets at
the roots~\cite{2310Noad}. 

To determine $k_\text{anchor}$, a known point in the stress-strain relationship other than $\sigma_{xx} = \varepsilon_{xx} = 0$ is required. Here, this is obtained using a feature
that appears clearly in both the stress-strain and the XRD data, a stress-induced first-order transition, and using the XRD data to obtain the absolute strain at this point. The
neutral strain point $\sigma_{xx} = \varepsilon_{xx} = 0$ was identified in sample B by an extremely small jump in displacement (not visible in any of the graphs to follow) that we
attribute to re-orientation of magnetic domains at zero stress~\cite{2507Stilkerich}. For sample C, no such feature was resolved, so the neutral strain point was instead located
using the force sensor reading after the sample had been fractured.

Although the two-spring approximation is a workable method of data analysis, it does introduce artefacts. At an ideal first-order structural transition, the Young's modulus
$d\sigma_{xx}/d\varepsilon_{xx}$ is expected to be zero across the strain range where phase coexistence occurs, and to jump discontinuously to nonzero values at the ends of this
range. We performed finite element analysis to see how such a transition would appear in our measurements. This simulation showed that the observed Young's modulus would be
zero over only about 70\% of the true strain range of phase coexistence, and that the discontinuous jumps would be smoothed into continuous ramps. Details are shown in the
Appendix.

\subsection{Neutron diffraction measurements}

The neutron diffraction measurements were performed during beamtime RB2320180 at the WISH instrument at the ISIS Neutron and Muon Source \cite{1100Chapon}. Two samples (D and E)
were measured. The samples were mounted into home-made uniaxial stress cells that incorporated strain gauges for sensing the applied displacement and force. Cadmium foil was used
to screen most of the cell from the neutron beam. The neutral strain point was identified from the force sensor reading after the sample fractured (sample D), or when a mechanical
contact built into the sample holder opened (sample E).

\section{Results: X-Ray Diffraction and Stress-Strain}
	
\subsection{X-ray diffraction data}
	
\begin{figure}[ptb]
\includegraphics[width=86mm]{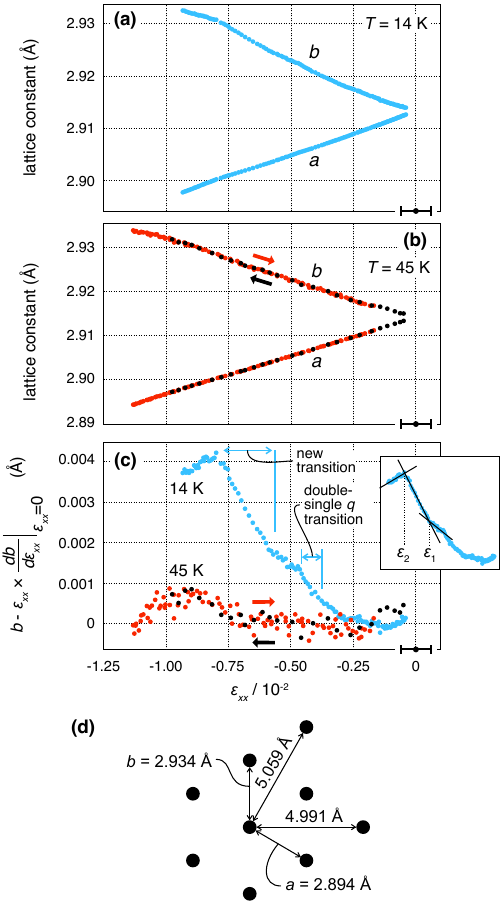}
\caption{\label{fig_XRDResults} Lattice constants $a$ and $b$ from XRD data, plotted against longitudinal strain $\varepsilon_{xx}$ (i.e. the strain along $[210]$). (a)
Lattice constants at $T = 14$~K.  (b) Same, at $T = 45$~K. (c) $b(\varepsilon_{xx})$ from panels (a) and (b), with background slopes subtracted.  The
inset shows the fits performed to identify the strain range $\varepsilon_1$ - $\varepsilon_2$ of the new transition. In all three panels,
the error bar on the $x$ axis is the error on $\varepsilon_{xx}$, $\pm 6 \cdot 10^{-4}$. (d) Nearest-neighbour and next-nearest-neighbour interatomic separations at the largest
strain achieved--- $\varepsilon_{xx} = -0.0113 \pm 0.006$ at 45~K. The errors on the given separations are approximately 0.001~\AA{}.}
	\end{figure}

The lattice constants $a$ and $b$ derived from XRD data, plotted against $\varepsilon_{xx}$, are shown in Fig.~\ref{fig_XRDResults}. Panel (a) shows data at 14~K, and panel (b) at
45~K. In these graphs, the error bar at the origin of the $x$ axis shows the error on $\varepsilon_{xx}$, $\pm 6 \cdot 10^{-4}$. This error is a consequence of systematic
uncertainty in $a$ and $b$: as was noted in the Methods section, measurement of sample A under strictly zero applied stress yielded $a = 2.9139 \pm 0.0014$~\AA{} and $b = 2.9126
\pm 0.0012$~\AA{} at $T = 14$~K, yielding an error of $\pm 0.002$~\AA{} on the quantity $a-b$, which in turn yields an error on $\varepsilon_{xx}$ of $\pm 6 \cdot 10^{-4}$. The
nearest-neighbour and next-nearest-neighbour atomic separations at the largest strain we achieved in the XRD measurements --- $\varepsilon_{xx} = -0.0113 \pm 0.006$ at 45~K --- are
shown in Fig.~\ref{fig_XRDResults}(d). 
	
To highlight nonlinearities in the dependence of $b$ on $\varepsilon_{xx}$, $b(\varepsilon_{xx})$ with a strain-linear background subtracted is shown in
Fig.~\ref{fig_XRDResults}(c). At 14~K, there are two regions, $-0.008 \lesssim \varepsilon_{xx} \lesssim -0.006$ and $-0.005 \lesssim \varepsilon_{xx} \lesssim -0.004$, where $b$
increases anomalously rapidly as the sample is compressed. The stress-strain data presented in the next section show that these regions correspond to first-order transitions.  The
transition at smaller compression is the double-single-$q$ transition, and the other is a new first-order transition. 

From the intersections of three linear fits to $b(\varepsilon_{xx})$, shown in the inset of Fig.~\ref{fig_XRDResults}(c), we determine that the new transition extends, at 14~K,
from strain $\varepsilon_1 = -0.59 \cdot 10^{-2}$ to $\varepsilon_2 = -0.80 \cdot 10^{-2}$--- a jump of $|\varepsilon_2 - \varepsilon_1| = 0.21 \cdot 10^{-2}$. The strain range
of the double-single-$q$ transition is not clearly resolvable in these data, but more clear in the stress-strain data presented below.

No peak splitting associated with the phase coexistence was resolved over the strain range $\varepsilon_2 < \varepsilon_{xx} < \varepsilon_1$. Our main result from the XRD
data is the differential Poisson ratio $\nu \equiv -d\varepsilon_{yy}/d\varepsilon_{xx}$, and this is shown together with the differential Young's modulus in
Fig.~\ref{fig_PoissonAndYoung} below.

	\subsection{Stress-strain relationship}

\begin{figure}[ptb]
\includegraphics[width=86mm]{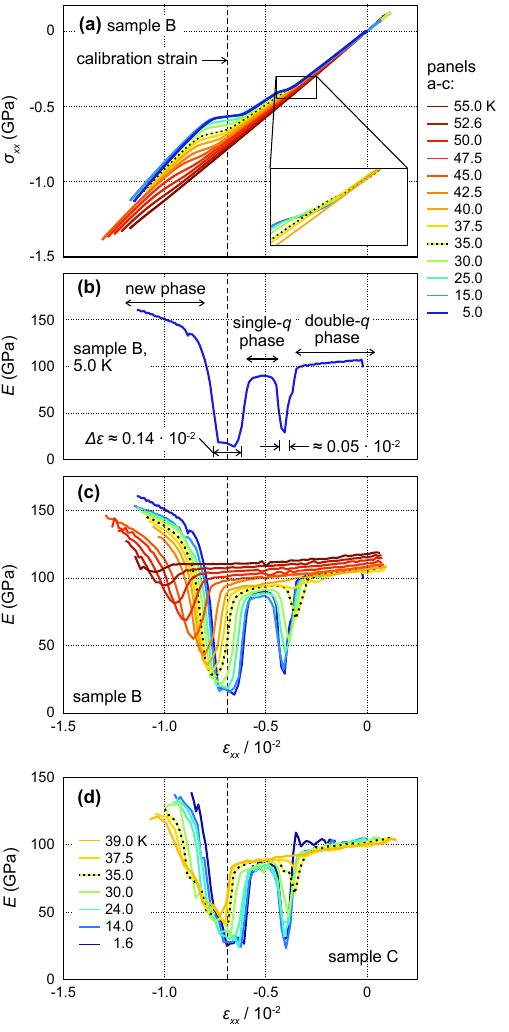}
\caption{\label{fig_stressStrain} Stress-strain relationship and differential Young's modulus. (a) Stress $\sigma_{xx}$ versus strain $\varepsilon_{xx}$ for sample B at
various temperatures. (b) Differential Young's modulus $E = d\sigma_{xx}/d\varepsilon_{xx}$ for sample B at 5.0~K, with the strain ranges of the three different phases
labelled.  (c) Differential Young's modulus of sample B at various temperatures. (d) Same for sample C. In all the panels, the calibration strain, $-0.69 \cdot
10^{-2}$, obtained from XRD data is shown by a vertical dashed line.}
	\end{figure}

Stress-strain data on sample B, obtained using the two-spring approximation (Eq.~\ref{eq_twoSpring}), are shown in Fig.~\ref{fig_stressStrain}(a). Immediately visible are two
strain jumps, which are the first-order transitions. The one at $\varepsilon_{xx} \approx -0.4 \cdot 10^{-2}$ is the double-single-$q$ transition, and the one at
$\varepsilon_{xx} \approx -0.7 \cdot 10^{-2}$ is the new transition.

As described in the Methods section, to determine the spring constant $k_\text{anchor}$ in the two-spring approximation we require a known fixed point in the stress-strain
relationship $\sigma_{xx}(\varepsilon_{xx})$ other than $\sigma_{xx} = \varepsilon_{xx} = 0$. We take for this point the midpoint of the new transition at $T = 15$~K, which was
found in the X-ray data to be at $\varepsilon_{xx} = (-0.69 \pm 0.06) \cdot 10^{-2}$. This strain is labelled ``calibration strain'' in
Fig.~\ref{fig_stressStrain}; $k_\text{anchor}$ in Eq.~\ref{eq_twoSpring} is set so as to place the midpoint of the new transition at this strain in the stress-strain data.  The
result is $k_\text{anchor} = 1.67 \pm 0.23$~N/$\mu$m for sample B, and $1.24 \pm 0.34$~N/$\mu$m for sample C.  The effects of the error on $k_\text{anchor}$ are discussed in the
Appendix.

In Fig.~\ref{fig_stressStrain}(a), it can be seen that the double-single-$q$ transition moves rightward slowly as $T$ is raised, and disappears at $T \approx 37.5$~K.  The new
transition, on the other hand, shifts leftward rapidly, to larger stress and strain, as $T$ is raised, and persists to temperatures above 50~K. The differential Young's modulus $E
\equiv d\sigma_{xx}/d\varepsilon_{xx}$ for sample B is shown in Figs.~\ref{fig_stressStrain}(b, c), and for sample C, in Fig.~\ref{fig_stressStrain}(d). The data from the two
samples reproduce each other well, though the data from sample B are cleaner at large strains. In the limit $T \rightarrow 0$, $E \approx 100$~GPa in the double-$q$ phase, $\approx
80$~GPa in the single-$q$ phase, and $\approx 150$~GPa in the new phase. Evidently, the magnetic order has a large effect on the Young's modulus. 

	\begin{figure}[ptb]
		\includegraphics[width=86mm]{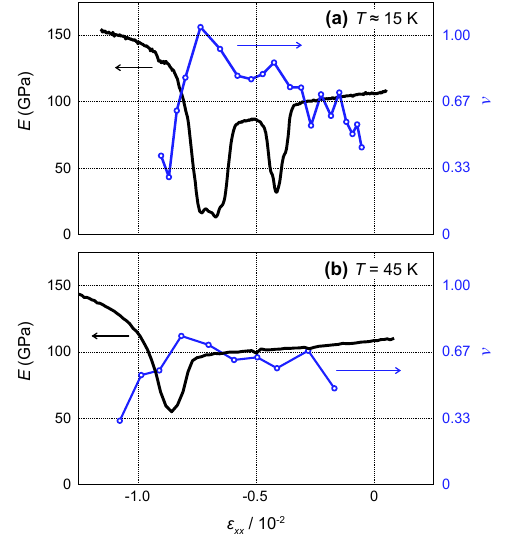}
		\caption{\label{fig_PoissonAndYoung} Differential Young's modulus and Poisson ratio. (a) Differential Young's modulus $E$ of sample B at 15~K, and
differential Poisson ratio $\nu$ of sample A at 14~K. (b) Equivalent data at 45~K.}
	\end{figure}

The strain jump at the new transition is, as indicated in Fig.~\ref{fig_stressStrain}(b), $0.14 \cdot 10^{-2}$ in the stress-strain data, whereas it was $0.21 \cdot 10^{-2}$ in the
XRD data. This difference is almost certainly an artefact of the two-spring approximation, mentioned above and described further in the Appendix, and the XRD result should be taken
as the correct one. At the double-single-$q$ transition, the strain jump obtained from stress-strain data is $\approx 0.05 \cdot 10^{-2}$, suggesting that the true strain jump at
this transition is $\approx 0.07 \cdot 10^{-2}$.  

The differential Young's modulus from sample B and differential Poisson ratio, $\nu \equiv -d\varepsilon_{yy}/d\varepsilon_{xx}$, from sample A are plotted together in
Fig.~\ref{fig_PoissonAndYoung}. At $T \approx 15$~K, the differential Poisson ratio is large: it is $\approx 0.5$ at zero stress, and $\approx 0.8$ just to the right of the new
first-order transition. It is $\approx 0.4$ in the new magnetic phase, which is a more typical value for a layered material~\cite{0205Paglione}. It is seen in
Fig.~\ref{fig_PoissonAndYoung}(b) that the behaviour at 45~K is similar, though not as pronounced: $\nu$ reaches $\approx 0.7$, then falls to $\approx 0.3$.

\begin{figure}[ptb]
\includegraphics[width=86mm]{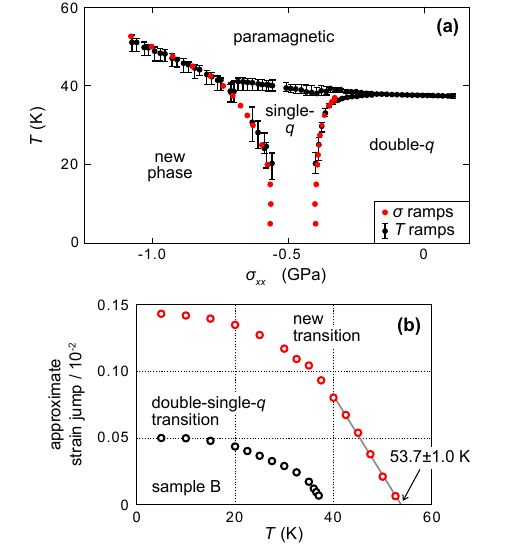}
\caption{\label{fig_phaseDiagram} Phase diagram. (a) Stress-temperature phase diagram of PdCrO$_2$, derived from stress-strain data on sample B. (b) Jump in strain
$\Delta \varepsilon$ at the two first-order transitions, as determined from stress-strain data on sample B. Due to artefacts introduced by the two-spring approximation, these
values for $\Delta \varepsilon$ are about 70\% of the true strain jumps at the transitions.}
	\end{figure}

A phase diagram derived from stress-strain data is shown in Fig.~\ref{fig_phaseDiagram}(a). The red points derive from the stress-ramp data shown above, while the black points
derive from temperature-ramp data (shown in the Appendix), where the transitions appear as anomalies in the thermal expansion. We do not analyse the sizes and shapes of these
anomalies, but the fact that the transitions appear in thermal expansion data shows that they are thermodynamic.

Finally, in Fig.~\ref{fig_phaseDiagram}(b) we show the magnitude of the strain jump across the two transitions, estimated using linear fits to the $\sigma_{xx}(\varepsilon_{xx})$
curves shown in Fig.~\ref{fig_stressStrain}(a). Although the strain jumps shown in this figure will be underestimates of the actual strain jumps (due to the limitations of the
two-spring approximation), the temperature dependence is useful information. The strain jump at the new transition appears to fall to zero at $T = 53.7 \pm 1.0$~K.

	\section{Results: Neutron Scattering}

The stress-strain and XRD data reveal a dramatic stiffening of the lattice across the new first-order transition: $E$ increases from $\approx 80$ to $\approx 150$~GPa, and $\nu$
falls from $\approx 0.8$ to $\approx 0.4$. In the Appendix, a model is presented of the equilibrium between the elastic and magnetic interaction energies, taking into account
nearest-neighbour magnetic interaction only. Using parameters suitable for PdCrO$_2$, it is shown that a spiral-N\'{e}el transition could yield changes in $E$ and $\nu$ of the
observed magnitude. One point from the analysis worth noting here is that the dependence of nearest-neighbour interaction energy on interatomic spacing might be even steeper than
has been thought. From comparison of different compounds with edge-sharing CrO$_6$ octahedra, this dependence has been estimated as $-40$~meV/\AA{}~\cite{7005Mochida, 1303Hardy},
but analysis of previously-published data on PdCrO$_2$~\cite{2112Sun} suggests an even steeper dependence: $-90 \pm 20$~meV/\AA{}.

To investigate the magnetic structure of the new phase, we turn to neutron diffraction.  It was shown in Fig.~\ref{fig_magOrderHypothesis}(e) how the magnetic Bragg reflections
are expected to move as the order evolves from a 120$^\circ$ spiral to N\'{e}el order. Pertinent to understanding the data that we now present: the reflection at $(h,k) =
(\frac{1}{3}, \frac{1}{3})$ is expected to move along a straight line to $(\frac{1}{2}, 0)$. (Since the applied strains are small on an absolute scale, we use the parent
R$\bar{3}$m unit cell to index magnetic reflections.)

\begin{figure*}[ptb]
		\includegraphics[width=\textwidth]{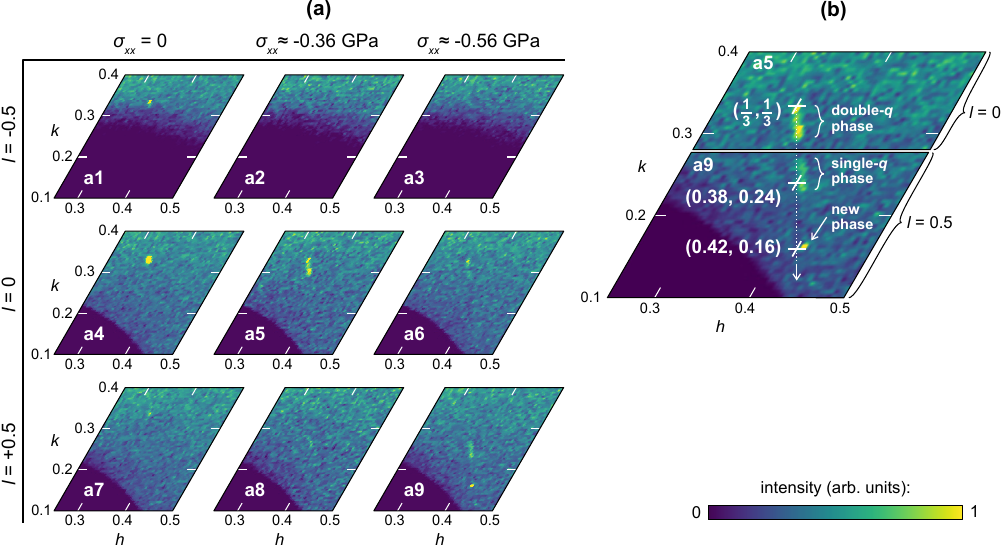}
		\caption{\label{fig_neutronScatteringData} Neutron scattering intensities from sample D at 8.5 K. (a) Intensities at three applied stresses, and for three
cuts along the $l$ axis.  All indexing is done with respect to the unstrained lattice vectors. (b) A collage of panels (a5) and (a9). The dotted line shows the expected
motion of the magnetic reflections for smooth evolution from the $120^\circ$ spiral to N\'{e}el order.}
	\end{figure*}

Neutron scattering intensities from sample D at three applied stresses are shown in Fig.~\ref{fig_neutronScatteringData}(a). Data were reduced and normalised using Mantid
v6.8~\cite{Mantid, 1411Arnold}. At each stress, three $hk$ cuts are shown, at $l=-0.5$, $l=0$, and $l=+0.5$. (For each cut, data were integrated over a range $\Delta l = 0.10$.)
At non-zero stress, the reflections appear as streaks rather than points. We attribute this to a combination of strain inhomogeneity and high sensitivity of the magnetic
periodicity to strain. The samples for neutron scattering were larger than for the XRD and stress-strain measurements, and larger samples of PdCrO$_2$ tend to have voids, around
which the stress will not be homogeneous. Additionally, neutrons are highly penetrating, and lower-stress regions of the samples were probed along with the exposed, high-stress
region.

Fig.~\ref{fig_neutronScatteringData}(b) is a collage of panels a5 ($\sigma_{xx} \approx -0.36$~GPa) and a9 ($\sigma_{xx} \approx -0.56$~GPa), and what is seen here is that the
scattering intensity tracks closely the expected line for evolution from $120^\circ$ to N\'{e}el order. The small deviation from this line could be due to misalignment between the
stress axis and $[210]$ direction.  

The intensity does not shift continuously. There is a streak in the $l=0$ plane, and then a gap across which the intensity jumps to the $l=+0.5$ plane. Both the gap and the jump to
a different $l$ value have been observed before~\cite{2112Sun}. They correspond to the double-single-$q$ transition, which is a first-order transition. The gap corresponds to a
range of propagation vectors that are not stable. The streak associated with the single-$q$ phase extends for a short distance, and then there is a large jump to a sharp
intensity peak at $(h,k) \approx (0.42, 0.16)$. We attribute this jump to the first-order transition into the new phase. 

\begin{figure}[ptb]
		\includegraphics[width=86mm]{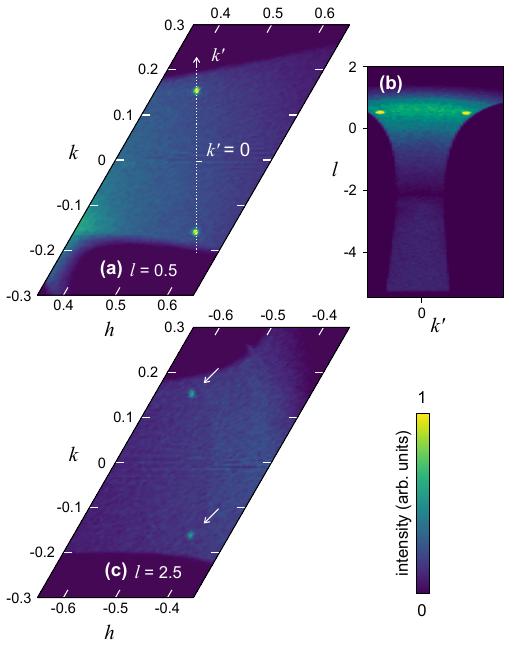}
		\caption{\label{fig_sampleE} Sample E at $T = 8.6$~K and $\sigma_{xx} \approx -0.72$~GPa. (a) A cut through the $l=0.5$ plane. (b) A cut along the
dashed line in panel a. N\'{e}el order would yield scattering at $k^\prime = 0$, corresponding to $(h,k) = (\frac{1}{2},0)$. (c) A cut through the $l=2.5$ plane.}
\end{figure}

The fact that the scattering intensity from the new phase is a sharp peak rather than a streak suggests that the new order is rigid, that is, responds weakly or not at all to
infinitesimal changes in lattice strain. That is consistent with the observed changes in elastic moduli, but $(h,k) = (0.42, 0.16)$ is close to the location where the $150^\circ$
spin spiral, illustrated in Fig.~\ref{fig_magOrderHypothesis}(c), would yield scattering intensity; the new phase is not N\'{e}el order. Shown in Fig.~\ref{fig_sampleE}(a,b)
are data from sample E at $\sigma_{xx} \approx -0.72$~GPa. Panel (a) shows a cut at $l = 0.5$, and although two reflections corresponding to the new phase are clearly visible,
there is no intensity above background at the N\'{e}el position. Panel (b) shows a cut along the $l$ axis, and it can be seen here that there is no intensity above the background
at the N\'{e}el position for any $l$ that we could observe. These observations raise an obvious question: what is special about the new propagation vector that makes it rigid?

The rigidity of the new phase might be a consequence of interaction with the Fermi surface of the Pd sheets. This hypothesis is taken up in the Discussion. For the remainder
of this section, let us look more closely at the propagation vector of the new stress-induced magnetic phase. To do so, it is necessary to consider domains. 

\begin{figure}[ptb]
		\includegraphics[width=86mm]{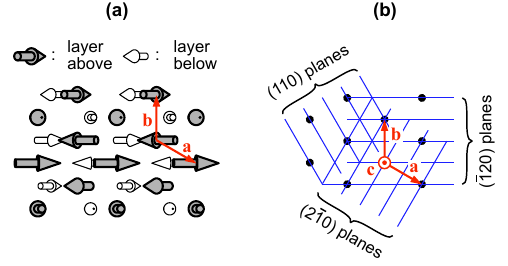}
		\caption{\label{fig_domainTypes} (a) Possible spin configuration of an obverse $(\bar{1}20)$ domain of the single-$q$ phase. This structure is obtained by
setting $\mathbf{q} = (0.375, -0.750, \frac{3}{2})$, and the spin planes to be the $(\bar{1}20)$ planes, as they are for $(\bar{1}20)$ domains of the double-$q$ phase in unstressed
PdCrO$_2$. (b) The $(\bar{1}20)$, $(2\bar{1}0)$, and $(110)$ planes--- the planes contain the indicated lines and the $c$ axis.}
\end{figure}

In Fig.~\ref{fig_neutronScatteringData}(b), we highlight a reflection that is within the single-$q$ phase, though close to the end of its range of stability: $(h,k) =
(0.38, 0.24)$. The three-fold rotational symmetry of unstressed PdCrO$_2$ yields three magnetic domain types, that we label $(\bar{1}20)$, $(2\bar{1}0)$, and $(110)$. Within the
single-$q$ phase, the propagation vectors of these domains are, respectively, $(q, -2q, \frac{3}{2})$, $(2q, -q, \frac{3}{2})$, and $(q,q,\frac{3}{2})$. According to the data of
Ref.~\cite{2112Sun} and Fig.~\ref{fig_neutronScatteringData}b, $0.367 < q < 0.382$. A possible spin configuration of $(\bar{1}20)$ domains of the single-$q$ phase is shown in
Fig.~\ref{fig_domainTypes}a. For this illustration, the spin planes are taken to be the $(\bar{1}20)$ planes, as they are for $(\bar{1}20)$ domains of the double-$q$ phase in
unstressed PdCrO$_2$.~\footnote{The magnetic structure of unstressed PdCrO$_2$ might not be co-planar, but it is at least close to being co-planar. Observed reflection intensities
are statistically indistinguishable from those of a co-planar structure~\cite{1403Takatsu, 1909Sun}.}

Conbining the three magnetic and two structural domain types, six domain types are possible in total. Each would yield intensity at $(h,k) = (0.38, 0.24)$, but at different
values of $l$. Previous measurements~\cite{2112Sun} have shown that, in the single-$q$ phase, uniaxial stress along $[210]$ favours $(\bar{1}20)$ domains.  For the obverse
structure type, intensity at $(h,k) = (0.38, 0.24)$ would appear as satellites of the $(0,1,3n-1)$ nuclear reflections, where $n$ is an integer.  Intensity would appear at
$l=\frac{1}{2}$, as observed. On the other hand, for the reverse structure type intensity at $(h,k) = (0.38,0.24)$ would be satellites of nuclear reflections at $(0,1,3n+1)$,
yielding intensity at $l = -\frac{1}{2}$, which was not observed. We conclude that the observed magnetic scattering intensity from sample D came dominantly from obverse structural
domains.

Extrapolating from the single-$q$ phase, we obtain a propagation vector for the new stress-induced phase of $\mathbf{q}_\text{new} \approx (0.42, -0.84, \frac{3}{2})$. Under
the applied uniaxial stress, the space group of the parent, non-magnetic cell changes from R$\bar{3}$m to C2/m, and this propagation vector is along the 2-fold axis. It is
consistent with all four observed magnetic reflections from the new phase. The number of observable reflections is limited by the access angle of the stress cell, and between
samples D and E, a total of four reflections were found that did not suffer strong interference from powder lines from the cell. The reflection at $(h,k) = (0.42,0.16)$ occurs at
$l = \frac{1}{2}$ for both samples D and E (see Fig.~\ref{fig_sampleE} for sample E), indicating that the magnetic intensity from sample E also came dominantly from obverse
structural domains. This reflection is a satellite of the nuclear reflection at $(0,1,-1)$.  The reflection at $(0.58, -0.16, \frac{1}{2})$, visible in Fig.~\ref{fig_sampleE}(a),
is a satellite of the $(1,-1,2)$ nuclear reflection.  Fig.~\ref{fig_sampleE}(c) shows a cut at $l = 2.5$, and the two reflections visible in this cut are satellites of the
$(0,-1,4)$ and $(-1,1,1)$ nuclear reflections.

Four reflections are not sufficient to attempt a refinement of the magnetic structure, but we can probe whether the new stress-induced phase is single- or double-$q$.  If it
were double-$q$, like the magnetic order of unstressed PdCrO$_2$, it would also contain the propagation vector $\mathbf{q}_2 = (0.42, -0.84, 0)$.  For both samples D and E, no
reflections were observed at locations where this propagation vector would yield intensity.  Two example cuts are shown in Fig.~\ref{fig_noDoubleQPeaks}, in the Appendix.  We
conclude that the new phase has a single propagation vector.

	\section{Discussion}

In the above, changes in the elastic moduli of PdCrO$_2$ under uniaxial stress, driven by stress-induced changes in the magnetic structure, were reported. There are two first-order
transitions: the double-single-$q$ transition and a transition newly reported here. The strain jump at the double-single-$q$ transition is $\Delta \varepsilon_{xx} \approx 0.07
\cdot 10^{-2}$, and at the new transition, $\approx 0.21 \cdot 10^{-2}$. At the new transition, the propagation vector jumps to $\mathbf{q}_\text{new} \approx (0.42, -0.84,
\frac{3}{2})$. From the sharpness of the reflections in the neutron scattering data in both Fig.~\ref{fig_neutronScatteringData}(a9) ($\sigma_{xx} \approx -0.56$~GPa) and
Fig.~\ref{fig_sampleE} ($\sigma_{xx} \approx -0.72$~GPa), and from the dramatic stiffening of the lattice, we conclude that the stress-induced magnetic order is rigid, meaning that
it changes minimally or not at all in response to infinitesimal changes in lattice strain. This rigidity is in sharp contrast to the magnetic orders at lower stresses, whose
propagation vectors shift rapidly as stress is applied. 

That the double-single-$q$ transition yields a strain jump of $\approx 0.07 \cdot 10^{-2}$ is potentially surprising, as this transition is a change in interlayer order, and
interlayer coupling is very weak in PdCrO$_2$. To look more deeply, a modified Helmholtz free energy may be defined: $F^\prime = U - TS + fL$, where $U$ is internal energy, $S$ is
entropy, $f$ is the force applied to the sample (with $f>0$ denoting tension), and $L$ is the sample length. The Young's modulus is given by $d^2 F^\prime/d\varepsilon_{xx}^2$, and
the change in $F^\prime$ associated with a strain jump $\Delta \varepsilon_{xx}$ may be estimated: 
\begin{equation}
\Delta F^\prime \approx \frac{1}{2}\left(\Delta \varepsilon_{xx}\right)^2 \frac{d^2F^\prime}{d\varepsilon_{xx}^2}.
\end{equation}
Taking $d^2F^\prime / d\varepsilon_{xx}^2 \sim 100$~GPa yields $\Delta F^\prime \sim 2 \cdot 10^4$~J/m$^3$, or $\sim 7 \cdot 10^{-6}$~eV per Cr atom. That is, on the single-$q$
side of the double-single-$q$ transition, $F^\prime$ of the single-$q$ configuration is about $7 \cdot 10^{-6}$~eV/Cr lower than that of the double-$q$ configuration. The scale of
this energy difference is consistent with experimental and theoretical findings that interlayer Cr-Cr interaction energies are well below 1~meV~\cite{1807Le, 2011Komleva}.

Let us now take up the question of why the new magnetic propagation vector is rigid. The new periodicity is close to a commensurate periodicity, $(h,k) = 
(\frac{5}{12},-\frac{5}{6})$, which would be obtained from the $150^\circ$ spin spiral illustrated in Fig.~\ref{fig_magOrderHypothesis}(c). But locking to such a high-order
commensurate periodicity is not expected, especially for a compound that has weak spin-orbit coupling. There is no detectable tendency of the magnetic order of PdCrO$_2$ to
lock to the lower-order commensurate position $(\frac{1}{3}, \frac{1}{3})$: the spontaneous rotational symmetry breaking in unstressed PdCrO$_2$ causes the magnetic periodicity to
differ from this commensurate position by around one part in a thousand~\cite{1909Sun, 2112Sun}.

\begin{figure}[ptb] \includegraphics[width=86mm]{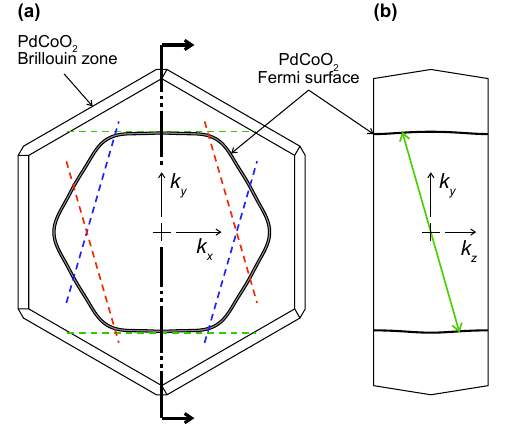} \caption{\label{fig_FSNesting}Illustration of nesting. (a) The Fermi surface of PdCoO$_2$ (a
non-magnetic analogue of PdCrO$_2$) plotted together with the magnetic Bragg planes expected from a spin spiral with $(h,k) = (0.42, -0.84)$. The colours correspond to the
real-space Bragg planes plotted in Fig.~\ref{fig_magOrderHypothesis}(c). The Fermi surface is obtained by summing the harmonic components reported in Ref.~\cite{1209Hicks}. (b) A
cross-section through panel a, illustrating the belly-neck nesting that would be obtained from $\mathbf{q}_\text{new} = (0.42, -0.84, \frac{3}{2})$.}
\end{figure}

The rigidity of the new phase might instead be a consequence of Fermi surface nesting. Shown in Fig.~\ref{fig_FSNesting}(a) is the Fermi surface of unstressed
PdCoO$_2$~\cite{1209Hicks, 0906Noh}, a non-magnetic analogue of PdCrO$_2$--- to very high accuracy, the Fermi surfaces of unstressed PdCrO$_2$ are a reconstruction of the PdCoO$_2$
Fermi surface~\cite{1507Hicks, 1401Noh, 1310Ok}. Also shown are the Bragg planes obtained from the propagation vector $\mathbf{q}_\text{new} = (0.42, -0.84, \frac{3}{2})$. The
green planes align well with the Fermi surface.  This propagation vector also provides belly-neck nesting, as as illustrated in Fig.~\ref{fig_FSNesting}(b). Although the
belly-neck distortion is small in PdCrO$_2$, the Fermi velocity is high, so belly-neck nesting is likely to be important in any nesting-driven phenomena.

If the new phase is indeed driven by nesting, then its rigidity under uniaxial stress indicates that the Fermi surface of the Pd sheets is much less sensitive to lattice strain
than the magnetism in the CrO$_2$ layers. The magnitude of the strain jump at the Fermi surface lock-in transition, $0.21 \cdot 10^{-2}$, corresponds to $\Delta F^\prime \sim 6
\cdot 10^{-5}$~eV per Cr atom. This is the energy that would be gained from Fermi surface nesting and the Cr-Pd coupling. A reasonable next step to test whether nesting really
is the key driver of the new magnetic phase is to explore theoretically whether such an energy gain is quantitatively reasonable.

\section{Conclusion}

The data above show that laboratory-achievable uniaxial stress can induce substantial changes in the magnetic structure of a frustrated magnet. The largest uniaxial stress
achieved here was $-1.4$~GPa, but for bulk oxides at least $-3$~GPa is possible~\cite{2208Jerzembeck}. The system that was studied here, PdCrO$_2$, can be considered a conceptually
simple starting point: classical, Heisenberg spins, and dominant nearest-neighbour interaction. It has been shown here that uniaxial stress applied along the $[210]$ lattice
direction induces two first-order magnetic transitions. The first is a change from a double- to a single-$q$ magnetic structure, and has been reported before. The second is a jump
to propagation vector $\mathbf{q}_\text{new} \approx (0.42, -0.84, \frac{3}{2})$, and is newly reported here. The propagation vector is much less sensitive to lattice strain after
this jump than before, and this rigidity is reflected in a dramatic stiffening of the lattice. The rigidity of the new propagation vector might be a consequence of Fermi surface
nesting.  Other magnetically frustrated compounds, such as candidate spin liquids, might be found in the future to show a similarly strong interplay between magnetic structure and
lattice elasticity.

	\section{Acknowledgements}

We thank Joseph Betouras and Frank Kruger for helpful discussions. NS, CWH and JG acknowledge funding from the Deutsche Forschungsgemeinschaft (DFG) (Project No. 429883427). NS,
JG, EG, and APM acknowledge funding from the Deutsche Forschungsgemeinschaft through SFB 1143 (Project No. 247310070, sub-projects C06 and C09). CWH acknowledges funding from the
Engineering and Physical Sciences Research Council (UK) (EP/X012158/1). EG, HMLN, and APM acknowledge funding from the Deutsche Forschungsgemeinschaft (TRR288-422213477
ELASTO-Q-MAT, sub-projects A10 and A13). JG acknowledges financial backing from the Würzburg-Dresden Cluster of Excellence on Complexity, Topology and Dynamics in Quantum Matter
(ctd.qmat EXC-2147, Project No. 390858490). NS, HMLN, KI, SK, EG, APM, and CWH acknowledge financial support from the Max Planck Society. 

	\section*{References}
	\bibliographystyle{unsrt}

	\section{Data Availability}
	
	The neutron scattering data are available at\\ \texttt{https://data.isis.stfc.ac.uk/doi/STUDY/}...\\ \texttt{120633327/}. The XRD and stress-strain data are available at\\
\texttt{https://doi.org/10.25500/edata.bham.00001577}.

	\section{Appendix}

	\subsection{Conversion from sensor units to force and displacement}

	To convert the raw force and displacement sensor capacitances to force and displacement, we apply the following relations~\cite{2310Noad}:
	\begin{eqnarray*}
		F = \nonumber \\
\epsilon_0 k_\text{f} A_\text{f} \left(\frac{1}{C_\text{f} - C_\text{f,offset}} - \frac{1}{C_\text{f,0} - C_\text{f,offset}}\right)  - k_\text{flex}D, \\
		D = \epsilon_0 A_\text{d} \left( \frac{1}{C_\text{d} - C_\text{d,offset}} - \frac{1}{C_\text{d,0} - C_\text{d,offset}}\right).
	\end{eqnarray*}
$k_\text{f}$ is the spring constant of the force sensor flexures, and $A_\text{f}$ and $A_\text{d}$ are respectively the force and displacement capacitor areas.  $\epsilon_0
k_\text{f} A_\text{f}$, $\epsilon_0 A_\text{d}$, $C_\text{f,offset}$, and $C_\text{d,offset}$ are all obtained from room-temperature calibration.  $\epsilon_0 k_\text{f}
A_\text{f}$ is multiplied by a factor of 1.15 as an estimate for the increase in the Young's modulus of the material of the stress cell, titanium, with cooling to cryogenic
temperatures~\cite{1902Barber}. $C_\text{f,0}$ and $C_\text{d,0}$ are the force and displacement capacitances at the zero stress point, which, as described in the Methods section,
was sometimes identified from features in the data, and other times by fracturing the sample. $k_\text{flex}$ is the spring constant of flexures in the sample carriers used for
some of the samples. These are functionally in parallel with the sample. $k_\text{flex} = 0.024$~N/$\mu$m for sample B, and 0.050~N/$\mu$m for sample C.

	\subsection{Artefacts introduced by the two-spring approximation}
	
	It was noted in the main text that the two-spring approximation, Eq.~\ref{eq_twoSpring}, introduces artefacts in reconstruction of the stress-strain relationship from
force-displacement data. To illustrate, finite element analysis was performed (using Comsol) of a fictional sample with a first-order transition. The sample is taken to have the
same dimensions as sample B. Its Young's modulus is set to $E = 110$~GPa for $\varepsilon_{xx} > -0.591 \cdot 10^{-2}$, zero for $-0.795 \cdot 10^{-2} < \varepsilon_{xx} < -0.591
\cdot 10^{-2}$, and 150~GPa for $\varepsilon_{xx} < -0.795 \cdot 10^{-2}$. Its differential Poisson ratio is set to 0.3 when $E \neq 0$, and 0.6 when $E=0$. For simplicity, the
elastic properties of this fictional sample are isotropic.

	\begin{figure}[ptb]
\includegraphics[width=86mm]{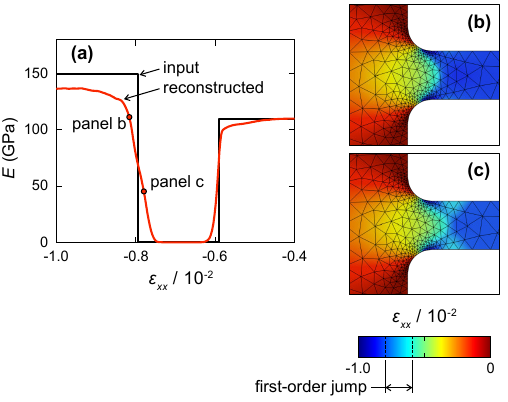}
\caption{\label{fig_strainSimulation} Artefacts introduced by the two-spring approximation. (a) Input Young's modulus of a fictional sample, and the reconstruction that
would be obtained were this sample measured as described above and analysed using the two-spring approximation (Eq.~\ref{eq_twoSpring}). (b), (c) $\varepsilon_{xx}$
around the root of the sample neck, at the labelled points in panel a.}
\end{figure}

	The black line in Fig.~\ref{fig_strainSimulation}(a) shows the assumed differential Young's modulus versus strain of this fictional sample, and the red line shows the
result that would be obtained if this sample were measured in our stress-strain apparatus and analysed using the two-spring approximation. Two artefacts are clearly visible. (1)
The strain range over which the reconstructed Young's modulus is zero is about 70\% of the range where the actual Young's modulus is zero. (2) The discontinuities at each end of
this range are broadened into ramps. These artefacts are consequences of strain inhomogeneity in the roots of the sample neck.  In principle, a more elaborate finite-element model
could be employed to extract the stress-strain relationship from force-displacement data, but due to geometrical uncertainties such an effort is unlikely to yield higher-confidence
results.

	\begin{figure}[ptb]
		\includegraphics[width=86mm]{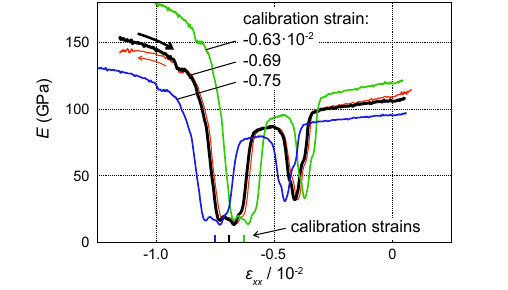}
		\caption{\label{fig_stressStrainError} Young's modulus $E$ versus strain $\varepsilon_{xx}$, for three possible values of the calibration strain, that is, the
strain at the midpoint of the new first-order transition.}
	\end{figure}
	
	\subsection{Effect of error on the calibration strain}

As was explained in the main text, a calibration strain --- a known point in the stress-strain relationship other than $\sigma_{xx} = \varepsilon_{xx} = 0$ --- is needed to extract
the stress-strain relationship from force-displacement data. The midpoint of the new first-order transition, $\frac{1}{2}(\varepsilon_1 + \varepsilon_2)$, where $\varepsilon_1$ and
$\varepsilon_2$ are the strains at either end of the transition, was selected as this calibration strain. From the XRD data, $\frac{1}{2}(\varepsilon_1 + \varepsilon_2) = (-0.69
\pm 0.06) \cdot 10^{-2}$. Shown in Fig.~\ref{fig_stressStrainError} is $E(\varepsilon_{xx})$ obtained with the calibration strain set to $-0.63$, $-0.69$, and $-0.75 \cdot
10^{-2}$. The error on the calibration strain translates into an error on the Young's modulus in the new magnetic phase of about $\pm 30$~GPa.

Also shown in Fig.~\ref{fig_stressStrainError} is $E(\varepsilon_{xx})$ obtained from an increasing-compression (right-to-left) and decreasing-compression (left-to-right) ramp. It
has previously been found that increasing-compression ramps are more likely to be affected by non-elastic deformation of the sample mounting epoxy~\cite{2211Barber}. This is seen
here, too: under high compression, the Young's modulus derived from the increasing-compression ramp is lower than that from the decreasing-compression ramp. All data shown in the
main text are from decreasing-compression ramps.

	\subsection{Temperature-ramp data}

\begin{figure}[ptb]
		\includegraphics[width=86mm]{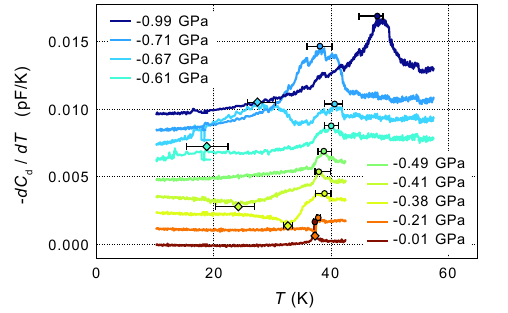}
		\caption{\label{fig_TRamps} Temperature-ramp data for sample B. $C_d$ is the differential change of the displacement capacitance with temperature when the voltage
applied to the piezoelectric actuators is held constant. The points and error bars correspond to transition temperatures identified in these data.}
	\end{figure}

	Shown in Fig.~\ref{fig_TRamps} is $-dC_\text{d}/dT$, where $C_\text{d}$ is the displacement sensor capacitance, for sample B. These data were collected with the voltage on
the piezoelectric actuators of the stress cell held constant, which kept the applied pressure approximately though not precisely constant. $C_\text{d}$ is inverse to displacement $D$,
so a peak in $-dC_\text{d}/dT$ corresponds to a peak in thermal expansion. Transitions were identified from peaks or dips in $-dC_\text{d}/dT$. The temperatures of the extrema were
taken to be the transition temperatures, while the width of the peak, judged by eye, was taken as the uncertainty on the transition temperature. These transition temperatures are
the black points plotted in Fig.~\ref{fig_phaseDiagram}(a).
	
\subsection{Additional Neutron Scattering Data}

\begin{figure}[ptb]
\includegraphics[width=85mm]{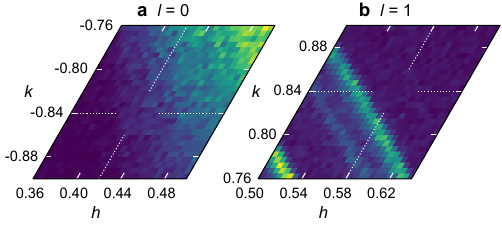}
\caption{\label{fig_noDoubleQPeaks} Cuts through \textbf{a} the $l=0$ plane and \textbf{b} the $l=1$ plane, for sample D under stress $\approx -0.56$~GPa. Intensity from the
propagation vector $\mathbf{q}_2 = (0.42, -0.84, 0)$ would appear at the intersections of the dotted lines.}
\end{figure}

Figs.~\ref{fig_noDoubleQPeaks}a,b show cuts of neutron scattering intensity through the $l=0$ and $l=1$ planes, respectively, for sample D. The dotted lines show where
reflections would have appeared were the new stress-induced phase a double-$q$ order, with propagation vector $\textbf{q}_2 = (0.42, -0.84, 0)$ along with
$\mathbf{q}_\text{new} = (0.42, -0.84, \frac{3}{2})$. A reflection in the $l=1$ plane would appear as a satellite of the $(1,0,1)$ nuclear reflection. For the obverse structure
type, nuclear reflections are permitted at combinations of $(0,-1,1)$, $(1,0,1)$, and $(-1,1,1)$.
	
	\subsection{Modeling} 

In this section, a calculation is presented of the elastic moduli of a triangular antiferromagnet as it is driven through a spiral-N\'{e}el transition by uniaxial stress, taking
into account nearest-neighbour magnetic interaction only. The new transition reported above might be a Fermi-surface lock-in transition, and to model such a transition
would require introducing Cr-Pd coupling, which we do not do. The goal here is simpler. The nearest-neighbour magnetic interation in PdCrO$_2$ is only $J \approx
6$~meV~\cite{1807Le}, and our goal is to see whether, with this value of $J$, transition to a rigid magnetic order could yield changes in elastic moduli of the observed magnitude.
We find that large changes in elastic moduli are indeed expected, due to the extreme sensitivity of $J$ to interatomic separation.

In Ref.~\cite{7005Mochida}, comparison of the N\'{e}el temperatures of different compounds with edge-sharing CrO$_6$ octahedra revealed a steep dependence of $J$ on interatomic
separation $a$: $dJ/da \approx -40$~meV/\AA{}. A very similar slope was derived from $T > T_\text{N}$ magnetic susceptibility data on different compounds~\cite{1303Hardy}. As we now
explain, data on PdCrO$_2$ under uniaxial stress suggest that $dJ/da$ might be about double this already-steep dependence. The analyses of Refs.~\cite{7005Mochida}
and~\cite{1303Hardy} are both strongly affected by LiCrO$_2$, which has exceptionally short Cr-Cr spacing, and therefore exerts strong influence on linear fits of $J(a)$. If
LiCrO$_2$ is excluded, larger $|dJ/da|$ is consistent with both data sets.

Under uniaxial stress, the magnetic component of the internal energy is
	\begin{equation}
		\label{eq_magInteractionEnergy}
		U_\text{mag} = \frac{1}{V}\sum\limits_{\langle ij \rangle} J_{ij} \mathbf{s_i} \cdot \mathbf{s_j} \\
		= \frac{2 J_1}{V} \cos \theta + \frac{J_2}{V} \cos (2\theta),
	\end{equation}
	where $\theta$ is the spin rotation angle across the $J_1$ bonds, $V$ is the volume of the unstressed lattice per Cr atom, and Eq.~\ref{eq_magInteractionEnergy} accounts
for nearest-neighbour interaction only. Fixing $dU_\text{mag}/d\theta = 0$ yields 
	\begin{equation}
		\theta = \arccos \left(-\frac{J_1}{2 J_2}\right).
	\end{equation}
N\'{e}el order --- $\theta = \pi$ --- is obtained for $J_2 \leq 0.5 J_1$.

In Ref.~\cite{2112Sun}, the $(h,k) = (\frac{1}{3}, \frac{1}{3})$ magnetic reflection was found to move under uniaxial stress at rate $dh/d\sigma = -0.047 \pm 0.009$~GPa$^{-1}$,
over a range of stress extending from zero to the double-single-$q$ transition. In the small-$\sigma$ regime, and assuming that $J$ depends on interatomic separation only,
$dh/d\sigma$ is given in the above magnetic model by 
\begin{equation}
\label{eq_rateOfChangeOfH}
\frac{dh}{d\sigma} = \frac{(1 + \nu)\sqrt{3}}{8\pi} \frac{1}{E} \frac{a_0}{J_0} \frac{dJ}{da},
\end{equation}
where $\nu$ is the Poisson ratio, $E$ the Young's modulus, $a_0$ the unstressed lattice constant, and $J_0$ is the nearest-neighbour magnetic interaction at zero stress. Averaging
over the range of stress between zero and the double-single-$q$ transition, we find here $E = 104 \pm 12$~GPa and $\nu = 0.61 \pm 0.03$. Setting $J_0 = 6$~meV,
Eq.~\ref{eq_rateOfChangeOfH} yields $dJ/da = -90 \pm 20$~meV/\AA{}. 

To evaluate the effect of the magnetic order on lattice strain, values of the elastic moduli without magnetic interactions are required. There do not appear to be published
experimental data on the elastic moduli of delafossite compounds, so we draw guidance from a different layered oxide, Sr$_2$RuO$_4$. At room temperature, its Young's modulus for
compression along the $[100]$ lattice direction is 175~GPa~\cite{0205Paglione}. For uniaxial stress applied along the $[100]$ direction, the in-plane Poisson ratio is
$-\varepsilon_{yy}/\varepsilon_{xx} = 0.394$, and the out-of-plane Poisson ratio is $-\varepsilon_{zz}/\varepsilon_{xx} = 0.207$. Following this guidance, we suppose an isotropic
Young's modulus of 180~GPa for PdCrO$_2$ in the absence of magnetic interactions, and set the nonmagnetic in-plane and out-of-plane Poisson ratios to 0.4 and 0.2, respectively.
Taking a tetragonal elastic tensor, these specifications yield elastic moduli in the Voigt notation of $C_{11,0} = 237.4$~GPa, $C_{33,0} = 207.7$~GPa, $C_{12,0} = 108.8$~GPa,
and $C_{13,0} = 69.2$~GPa, where the subscript 0 denote that these are the moduli without magnetic interactions. The elastic energy is given by
	\begin{equation}
		U_\text{el} = \frac{1}{2}\hat{\varepsilon} \hat{C_0} \hat{\varepsilon}^T,
	\end{equation}
	where $\hat{\varepsilon} = (\varepsilon_{xx}, \varepsilon_{yy}, \varepsilon_{zz})$, and 
	\begin{eqnarray}
		\hat{C_0} = \left(\matrix{
			C_{11,0} & C_{12,0} & C_{13,0} \cr
			C_{12,0} & C_{11,0} & C_{13,0} \cr
			C_{13,0} & C_{13,0} & C_{33,0}} \right).
	\end{eqnarray}
	The $x$, $y$, and $z$ directions are, respectively, the $[210]$, $[010]$, and $[001]$ lattice directions (see Fig.~\ref{fig_magOrderHypothesis}).
	
	\begin{figure}[ptb]
		\includegraphics[width=86mm]{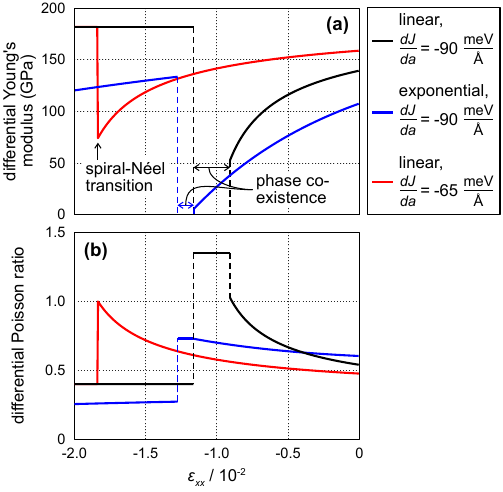}
		\caption{\label{fig_model}Results from a baseline model for elastic deformation across the spiral-N\'{e}el transition. (a) Differential Young's modulus $E$, and (b)
Differential Poisson ratio $\nu$. The nearest-neighbour interaction in unstressed PdCrO$_2$ is taken to be $J_0 = 6$~meV. Results are shown for three different assumptions of the
dependence of $J$ on interatomic separation $a$. For the exponential dependence, $dJ/da$ is set to $-90$~meV/\AA{} at $a = a_0$, the interatomic separation in unstressed PdCrO$_2$.}
	\end{figure}

Our simulation is performed at $T \rightarrow 0$, so the Helmholtz free energy is $F = U_\text{el} + U_\text{mag}$ (with no entropy term), and $E = d^2(U_\text{el} +
U_\text{mag})/d\varepsilon_{xx}^2$. Simulation parameters are set to yield $J_0 = 6$~meV and $a_0 = 2.9132$~\AA{} when the lattice elasticity and magnetic interactions with
$120^\circ$ spiral order are in equilibrium, without applied stress. Taking $\varepsilon_{xx}$ as the independent variable, $\varepsilon_{yy}$ and $\varepsilon_{zz}$ are then
calculated to minimise $U_\text{el} + U_\text{mag}$.

Results are shown in Fig.~\ref{fig_model}, for three examples of $J(a)$. The first is a linear dependence of $J$ on $a$, with $dJ/da = -90$~meV/\AA{}. The second is an exponential
dependence, with $dJ/da = -90$~meV/\AA{} at $a = a_0$. The third is a linear dependence with $dJ/da = -65$~meV/\AA{}. It can be seen in the results that the spiral-N\'{e}el
transition is continuous for smaller $|dJ/dai|$, and first-order for larger $|dJ/da|$. The ranges of phase coexistence for the first-order transitions are calculated assuming
negligible domain-wall energy. Although, given the strain mismatch, the domain wall energy is likely to be quite high, the long-and-narrow shape of our samples will reduce the
energy cost of domain walls~\cite{0909Cao}.

Very different results are obtained for a linear versus an exponential dependence of $J$ on $a$; in any attempt to accurately reproduce the behaviour of PdCrO$_2$, the
dependence of $J$ on $a$ would need to be specified carefully. The more basic point is amply demonstrated by the simulation results: variation of the Young's modulus by $\sim
100$~GPa, and a Poisson ratio reaching or even exceeding 1, are reasonable outcomes from the known strength of magnetic interaction in PdCrO$_2$ and its observed sensitivity to
interatomic spacing.

\end{document}